\documentclass[SingleSpace,NoLineNumbers, InsideFigs]{asce_style}
\usepackage[utf8]{inputenc}
\usepackage[T1]{fontenc}
\usepackage{lmodern}
\usepackage{graphicx}
\usepackage[figurename=Fig.,labelfont=bf,labelsep=period]{caption}
\usepackage{subcaption}
\usepackage{amsmath}
\usepackage{enumitem}
\usepackage{natbib}
\usepackage{newtxtext,newtxmath}
\usepackage{hyperref}
\hypersetup{colorlinks=true,citecolor=blue,linkcolor=black}

\usepackage{newunicodechar}
\newunicodechar{ﬁ}{fi}
\newunicodechar{ﬀ}{ff}
\NameTag{Rahbaralam, \today}
\begin{document}

\title{Predictive Analytics for Water Asset Management: Machine Learning and Survival Analysis }

\author[1,4]{M. Rahbaralam}
\author[1]{D. Modesto}
\author[2]{J. Card\'{u}s}
\author[3]{A. Abdollahi}
\author[1]{F.M. Cucchietti}

\affil[1]{Barcelona Supercomputing Center (BSC), C/Jordi Girona 29, 08034, Barcelona, Spain}
\affil[2]{Aig\"{u}es de Barcelona, General Batet 1‐7, 08028 Barcelona, Spain}
\affil[3]{Laboratori de C\`{a}lcul Num\`{e}ric (LaC\`{a}N), Universitat Polit\`{e}cnica de Catalunya (UPC), Campus Nord UPC-C2, E-08034 Barcelona, Spain}
\affil[4]{Corresponding author: maryam.rahbaralam@bsc.es}

\maketitle

\begin{abstract}
Understanding performance and prioritizing resources for the maintenance of the drinking-water pipe network throughout its life-cycle is a key part of water asset management. Renovation of this vital network is generally hindered by the difficulty or impossibility to gain physical access to the pipes. We study a statistical and machine learning framework for prediction of water pipe failures. We employ classical and modern classifiers for a short-term prediction and survival analysis to provide a broader perspective and long-term forecast, usually needed for the economic analysis of the renovation. To enrich these models, we introduce new predictors based on water distribution domain knowledge and employ a modern oversampling technique to remedy the high imbalance coming from the few failures observed each year. For our case study, we use a dataset containing the failure records of all pipes within the water distribution network in Barcelona, Spain. The results shed light on the effect of important risk factors, such as pipe geometry, age, material, and soil cover among others, and can help utility managers conduct more informed predictive maintenance tasks.

\end{abstract}

\section{Introduction}

The drinking-water transport and distribution network is one of the most valuable assets of a water utility company, as well as a critical urban infrastructure. A considerable amount of effort is devoted to optimally plan the annual preventive replacements of pipes and other elements \citep{MAHMOODIAN201849}, as proper network maintenance can lead to significant benefits, better understanding of the assets, and improvement of business processes and services in a sustainable manner. Physical assets, such as pipes, tend to deteriorate over time, and aging drinking water pipes face an increasing number of failures leading to service interruptions, higher operating costs, and water and energy loss \citep{Berardi2008}. Deteriorating water pipes threaten a water utility’s ability to deliver safe and secure drinking water to its customers, consequently affecting the city life and the environment \citep{HUKKA2015112}.

Among the many steps involved in asset management, developing an accurate model to predict remaining assets' lifespan is a key component. Predicting pipe failures and identifying the vital factors play a significant role in strategic rehabilitation planning of water distribution infrastructures \citep{Syachrani2013}.  A wide range of pipe-break prediction models has been developed over the last decades to aid utility companies as they develop their long-term asset management plans and improve their water security. The most common way to predict pipe failure is to apply statistical models based on analysing various levels of historical data \citep{Shahata}.
\citet{SCHEIDEGGER2015237} provided an overview of statistical models and \cite{Kleiner}; \cite{Martins2013} and \cite{Osman2011} compared the strengths, weaknesses, and limitations of these models. A classical statistical approach is survival analysis, predicting the survival probability of pipes over time. One of the most famous regression techniques for survival analysis is the Cox Proportional Hazards Model (PHM), which has been used to relate several risk factors to survival time \citep{Bagdonavicius}. The potential of survival analysis relies on its ability to provide a long-term prediction of pipe failure probability \citep{Lee}. 

Recently, artificial intelligence and machine learning (ML) models have provided a new management tool to produce a more accurate prediction of water pipe probability of failure. The task at hand is typical of classification models: grouping the data into predefined classes and predict the class to which the previously unseen data belongs \citep{HEUNG201662}. When the goal of a classification model is to understand the underlying relationships among predictors and target variable (failure), Logistic Regression (LR) analysis is one of the best methods available, providing an easily interpretable effect of each predictor \citep{Yamijala}. Recently developed models have been reported to predict pipe failures in water supply networks using LR \citep{ROBLESVELASCO2020106754}. On the other hand, there are other approaches capable of capturing nonlinear patterns in the data, sometimes at the price of interpretability. \cite{Achim2007} applied Neural Networks to predict pipe failure in a two year horizon. \cite{Shirzad} compared the performance of Artificial Neural Network (ANN) and Support Vector Regression (SVR) machine learning algorithms for predicting the pipe failure rate in water distribution networks. However, to function properly, ANN models commonly need large datasets, usually much larger than what typical digital histories of utility companies. Furthermore, the main limitation of SVR models is their considerable computational cost with quadratic complexity in the number of samples. Decision tree learning techniques are also popular models for classification, and thus for pipe failure prediction. \cite{Syachrani} developed a decision tree-based deterioration model for sewer pipes. Some other applications of decision-tree models include those developed by \cite{Harvey2014}, \cite{Sun2014} and \cite{Winkler2018}. Among the decision-tree methods, eXtreme Gradient Boosting (XGBoost) has gained much popularity as the chosen algorithm of many winning teams in machine learning competitions in Kaggle \citep{Tianqi2016}, an online community of data scientists and machine learners. Although these sophisticated models are able to discover complex patterns in the data, it can be difficult to extract from them an interpretable and consistent effect of predictors as offered by LR models. Surrogate models have been recently developed to enable better interpretability of decision-tree models, among which, SHapley Additive exPlanations (SHAP) is one of the most consistent approaches \citep{Lundberg2018}.

The number of annual pipe failures in a water distribution network is typically small compared to the total number of installed pipes. Therefore, there is a strong imbalance or uneven distribution of classes between the rare or minority category (failure) and the common or majority category (no failure) \citep{NISBET200949}. Not all the above-mentioned ML methods are immune to the effect of class imbalance--with the main impact being that the results might be biased towards the common class, diminishing the effectiveness of the whole approach \citep{Krawczyk}. To the best of our knowledge, the majority of pipe failure models developed so far for water distribution systems have not dealt with this class imbalance. One practical approach is to re-balance the target (failure) by calculating pipe failures for a longer period, consequently increasing the total number of pipes in the rare class. More traditional solutions are based on resampling (over- or under-sampling) the dataset to achieve a better class balance. Among these techniques, the Synthetic Minority Over-sampling Technique (SMOTE) is a powerful approach that creates synthetic minority class samples based on nearest neighbours \citep{Chawla02smote}. This approach has proven successful in a variety of applications from several different domains \citep{WAHYUPRATAMA2018174}. In addition to these balancing techniques, proper evaluation metrics should also be considered in dealing with imbalanced datasets, since standard evaluation metrics for classifiers, e.g. accuracy, treat all classes as equally important, leading to a biased evaluation towards the majority class. The Matthews Correlation Coefficient (MCC), instead, is a more reliable metric which considers the ratio between the sizes of the minority and majority classes in an imbalanced dataset \citep{Chicco2020}. 

Another fundamental step for the application of ML methods is to identify input features or predictors, which enclose useful information for predictions. Feature engineering is the process of discovering new variables based on domain knowledge. Although most ML models are able to extract useful information from existing or standard features, they cannot identify nonlinear combinations of the features. Good predictors incorporate knowledge that is foreign to the original dataset, consequently enhancing the performance of ML models to detect pipe failures.

To overcome the limitations of the existing approaches, in this study we introduce a framework for the annual prediction of pipe failures in water distribution networks by exploiting and combining the strengths of the above-mentioned statistical and ML methods. We develop this framework using data from the Aigues de Barcelona (AB) water utility company, which contains failure records of the last 16 years for each pipe within the water distribution network in Barcelona. AB's drinking water supply network consists of about 4,500 km of pipelines, mostly running underground. Unexpected failures in the network can potentially interrupt the water supply to a significant part of the three million inhabitants serviced by AB, including the service to 89 medical centers sensitive to water service, with important impacts to the city life and the environment. The main goal of our proposed framework is to maximize the prediction of these unexpected failures and to flag critical pipes for rehabilitation.

The structure of this paper is as follows: In Section 2, we present the data pre-processing process, including data gathering and cleaning, feature selection, feature engineering, and dataset balancing. The modelling procedure is introduced in Section 3, where we employ Logistic and XGBoost classification models to predict short-term pipe failures, and survival analysis to provide a long-term prediction of failure probability. We express conclusions and future directions in the last section.

\section{Data Preprocessing}
\label{DataPre}
\subsection{Data gathering and cleaning}


A large number of raw data, including inventories of pipe-sections of the water distribution network, were recorded annually by AB from 2002 to 2019. These datasets are enriched with additional features which could have a potential correlation with pipe failures. Some of these pre-processed features cannot be obtained directly but are derived from compositions of other features. For many raw data features it is necessary to design and employ a specific validation and correction process. The long history of service by AB, and more importantly, the considerable expansion of the network to different areas explains the variability of the quality of the data. We do not consider the datasets of 2002 and 2003 due to a comparatively higher percentage of missing values in these datasets. The data pre-processing is performed with the KNIME data analytics platform \cite{KNIME}. 

There is an alphanumeric inventory of the pipe-sections of the network for each year since 2001. This inventory was obtained on December 31 of each year. The structure of datasets is different before and after 2009; therefore, it is necessary to modify and reconstruct the missing fields using other sources of information. For example, the feature $Material$ does not follow the same naming convention between years and a consistent mapping is implemented for this field for all the years. Also, around 9$\%$ of the pipes have no installation date value. An approximate date was obtained using related information like the material, geographic information, and the network development in the area and neighboring ones. 
Apart from the date of installation, we also obtain the age of pipes connected to any one pipe and compute its average, and the same for the age of accessories and elements connected to a pipe.

Apart from the inventory datasets, there are also two datasets for failures and geolocated information of the pipes. The former contains all the details including repair procedures, alerts, and costs. From the geospatial dataset we obtain information such as cartographic and repair operations locations. The merging of these datasets and the inventory is done through the pipe ID. 

Geolocation of the pipes is then crossed with terrain information to infer the length of each pipe that is under a sidewalk, under streets, or under uncovered soil (e.g. parks, planted areas, etc.).

Finally, we perform a hydraulic pressure simulation of the network using the daily pressure data of pipe nodes. This simulation allows us to estimate the stress and pressure amplitude for each pipe in different years, using the yearly evolution of district metering areas and piezometric zones. The difference between the maximum and minimum estimated pressures was calculated from the simulation results, while average annual pressures are obtained from the pipe end point altimetries and its piezometric head.


\subsection{Feature Selection}

The initial data pre-processing leads to 16 yearly datasets, from 2004 to 2019, with 44 features containing information of around 120,000 water pipes installed in the province of Barcelona, Spain. In the first step, we concatenate these datasets to a single dataset and remove features that do not have any technical relevance for the model, including additional individual labels. Finally, 21 features are selected and summarized in Table \ref{tablefeatures}. Note that the inclusion of any of the remaining 23 features does not enhance the performance of the ML models discussed in the next section. The variable of interest or target feature is \textit{Failure}, which is a binary feature indicating whether each pipe has had any failure (class = 1) or not (class = 0) in a given year. There are 20 predictors, excluding the target feature and the pipe ID, among which 12 are numeric, and 8 are categorical features.

\begin{table}[h!]  
\begin{center}
\caption{Features in the Final dataset} 
\label{tablefeatures}
\centering 
\begin{tabular}{p{40mm}|p{70mm}|p{20mm}} 
\hline\hline   
 Feature & Description  & Type
\\ [0.5ex]  
\hline   
ID & Pipe unique identifier & Numeric \\ 
Material & Pipe material & Categorical\\
Diameter  & Pipe diameter & Numeric\\
Length  & Pipe length & Numeric\\
Failure  & Pipe failure & Categorical\\
Install year  & Pipe installation year & Numeric\\
Num-connections & Number of connected pipes & Numeric \\
Avg-connections & Average age of the connected pipes & Numeric \\
Num-elements & Number of elements and devices in the pipe & Numeric \\
Avg-elements & Average age of the elements & Numeric \\
City  & Pipe location (city/zone) & Categorical   \\ 
Network type & Function of the pipe (distribution or transport) & Categorical\\

Sidewalk Length & Length of section that is under sidewalk & Numeric  \\
Ind-greenzone & Pipe indicator by green zone & Categorical   \\ Green zone length  & Length of section that is under green zone & Numeric   \\
Assimilable to transport  & Pipe operating conditions & Categorical\\
Level of traffic & Level of traffic intensity& Categorical\\
Underground water gallery & Passing through underground water gallery & Categorical\\ 
Pressure & Average hydraulic pressure of the pipe  & Numeric  \\
MaxvsMin-Pressure & Difference between the maximum and the minimum simulated pressures & Numeric \\
Estres-Pressure & Sum of the simulated pressure rises & Numeric \\
\hline 
\end{tabular}
\end{center}
\end{table}

\subsection{Feature Engineering}
\label{FeatEng}
Categorical features must be encoded into numerical labels, because many ML algorithms can only read numerical values. We apply a one-hot encoding to categorical features such as \textit{Materials}, \textit{City} and \textit{Network type}, by adding a new binary variable for each category and removing the original categorical features. Moreover, in several ML algorithms based on stochastic gradient descent optimization, feature scaling can noticeably improve the convergence speed of the algorithm \citep{Buntine}. Here, we employ feature standardization which makes the values of each feature in the data have zero-mean and unit-variance \citep{Grus:2015}.

We add some simple new features like the age of each pipe (\textit{Age}), calculated by subtracting the installation date from the record date, rounding to years. We calculate the total number of failures for each pipe up to a given year and assign it to the new feature \textit{Failure history}. This task includes tracking all pipes to their origin to account all their previous failures, because, after a failure, some pipes were split and assigned a new pipe ID. The operations performed to repair a pipe may also affect other adjacent pipes. To account for this effect, the total number of annual repairing operations in the vicinity of each pipe (< 2 meters) is calculated as the new feature \textit{Pipe operations}. A similar feature, \textit{Accessory operations}, is also created for the operations on accessories in the vicinity of each pipe. Other forms of operations --e.g. installation of measuring equipment or neighboring pipes, anything that requires shutting off the network in the area -- are also counted and assigned to the new feature \textit{Other operations}.

We also introduce some non-linear combinations of the existing features, such as the ratio of length under a sidewalk or under green zone to the total length of the pipe, and the ratio between the age of pipe connections or elements to the pipe's age. Two features are of notice: one is \textit{LongRatio}, the ratio between the current and the original length of the pipe, which takes into account that over time some pipes are split or cut due to failures or for expanding or changing the distribution network. The second one is \textit{Aspect ratio}, we
define as the ratio of \textit{Diameter} to \textit{Length}, although it is not a purely geometric feature as the length of some pipes is computed as the sum of lengths of many pipes (same material, diameter, and installation date) connected in series. Table \ref{tableNew} presents all the new features.


\begin{table}[!ht]
\renewcommand{\arraystretch}{1.50}
\caption{New Features}
\label{tableNew}
\centering
\begin{tabular}{p{70mm}|p{90mm}}
\hline
\bfseries Feature & \bfseries Explanation \\
\hline
Age & ${\text {Pipe age}}$  \\
Failure history & Accumulated numbers of failures \\
Aspect ratio & Diameter of the pipe / Length of the pipe  \\
Sidewalk length ratio   & Sidewalk length / Length of the pipe\\
Green zone length ratio &   Green zone length / Length of the pipe  \\
Age-connections ratio &  Average age of connected pipe / Age of the pipe  \\
LongRatio & Current length / Original length \\
Pipe operations & Number of operations due to pipe failure, marked in the geographical system and less than 2 meters from the pipe\\
Accessory operations & Number of operations due to breakage of section accessory, marked in the geographical system and less than 2 meters from the pipe\\
Other operations &  Number of other operations marked in the geographical system and less than 2 meters from the pipe\\
\hline
\end{tabular}
\end{table}

\subsection{Dataset Balancing}

Around 99\% of data belongs to the negative class 0 and around 1\% to the positive class, highlighting that our dataset is highly imbalanced. Imbalanced datasets are a special case for classification problems where the class distribution is not uniform among the classes. Typically, they are composed of two classes: The majority (negative = 0) class and the minority (positive = 1) class. The difficulty with imbalanced datasets is that classification methods are often biased towards the majority class. Considering our dataset, it means that if classifiers always predict no failure, this prediction would be correct in \(99 \%\) of the cases.

One way to mitigate this bias is to re-balance the target feature by considering pipe failures for a longer period. For instance, instead of one year, if the target feature $\textit{Failure}$ represents the pipe failures in four years, the number of the positive cases 1 (failure) increases to around 4\%. Note that, to avoid data leakage between the training and validation sets, this increase results in a larger gap between these datasets, consequently decreasing the number of training samples and cross-validation folds. However, as we show in the next section, this re-balancing method leads, overall, to higher classification scores obtained by ML models.\\         
Another effective method to deal with imbalanced datasets is over-sampling. \cite{Chawla02smote} developed the Synthetic Minority Over-sampling Technique (SMOTE), which synthetically generates new minority class samples in the feature space rather than in the data space to balance the class distribution and widen the decision region of the minority class. In SMOTE, a new synthetic minority class sample is generated, which lies on the line segment between the minority class sample $x_{i}$, which is to be oversampled, and another minority sample $\bar{x}$, which is usually selected from the $N_{\text {min }}$ samples near $x_{i}$. The synthetic sample is then obtained as  

\begin{equation}
x_{\mathrm{syn}}=x_{i}+\left(\bar{x}-x_{i}\right) \cdot \operatorname{rand}(0,1),
\end{equation}
where rand(0, 1) indicates a random number within the interval \((0,1)\). This technique has proven its potential to increase the performance of ML algorithms \citep{Alberto2018}. After employing the SMOTE technique on our dataset the two classes have an even distribution.

\section{Classification}
\label{ClassSec}
Classification is a supervised learning method for making discrete predictions, given unforeseen input instances. In this study, we employ classification to predict whether or not a pipe failure occurs in a short time (1 to 4 years). The variable of interest or target feature is \textit{Failure}, which takes bindery values 0 (no failure) or 1 (failure) if the pipe has a failure in the following years. ML classifiers commonly output a probability of failure $p$ for each pipe, and a binary (discrete) output can be obtained by considering the classification threshold (also called the decision threshold) as  

 \begin{equation}
 y_{i}=\left\{\begin{array}{l}
 {0 \text { if } p_{i} \leq \text { threshold }} \\
 {1 \text { if } p_{i}>\text { threshold }}
 \end{array}\right.
 \end{equation}
where $y$ in the binary output and the subscript $i$ represents the pipe ID. The classification threshold is one of the parameters which should be tuned to obtain the best performance of ML classifiers with respect to the desired metric of choice.

\subsection{Evaluation Metrics}

Among the standard performance evaluation scores for classifiers, the Matthews Correlation Coefficient (MCC) \citep{MCC1975} is one of the metrics which correctly takes into account the ratio of the confusion matrix size, and it contains the balance ratios of the four confusion matrix categories, true positives (TP), true negatives (TN), false positives (FP), and false negatives (FN) as

\begin{equation}
\mathrm{MCC}=\frac{T P \times T N-F P \times F N}{\sqrt{(T P+F P)(T P+F N)(T N+F P)(T N+F N)}}.
\end{equation}
Especially on imbalanced datasets, MCC is  able to correctly inform whether prediction evaluation is going well or not, while accuracy or F1-score would not \citep{Chicco2017, Chicco2020}. In addition to MCC,  our evaluation metric, we also obtain the precision and recall (useful for interpretability and to accomodate business criteria) as  
\begin{equation}
\mathrm{Precision}=\frac{T P }{(T P+F P)},
\end{equation}

\begin{equation}
\mathrm{Recall}=\frac{T P }{(T P+F N)}.
\end{equation}
These scores are computed as a function of the classification threshold using the following cross-validation procedure. 

\subsection{Validation and Results}
In the case of our four-year prediction, we aim to predict pipes failure from 2016 to 2019 given the data vailable until 2015. To provide an unbiased evaluation of ML methods, we first separate the dataset of 2015 as our test dataset and the data between 2004 to 2011 as the training set. The four-year gap between these datasets is considered to avoid data leakage to the test set.

\begin{figure}[!htp]
  \centering
  \includegraphics[width=0.78\textwidth]{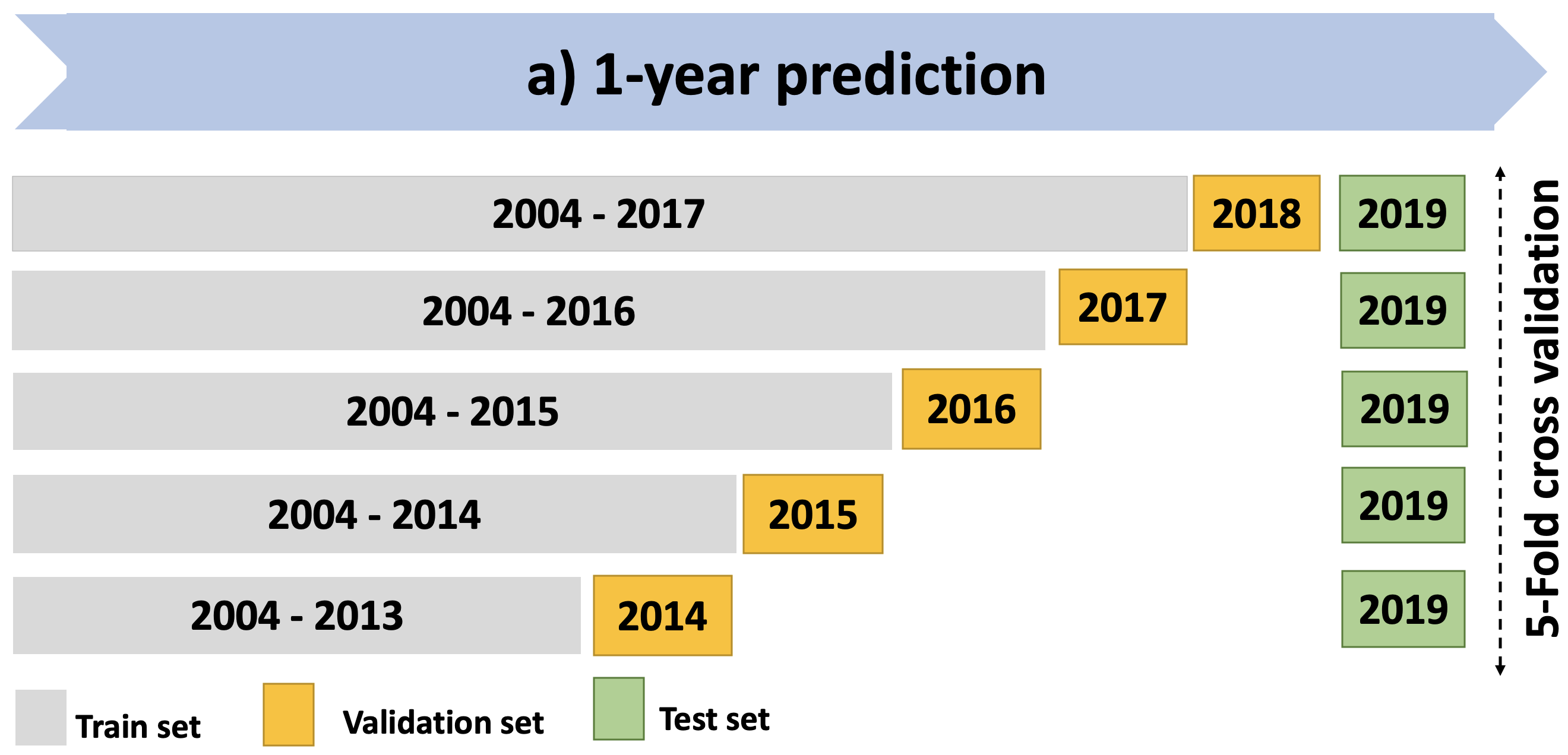}\\
   \vspace{1cm}
  \includegraphics[width=0.78\textwidth]{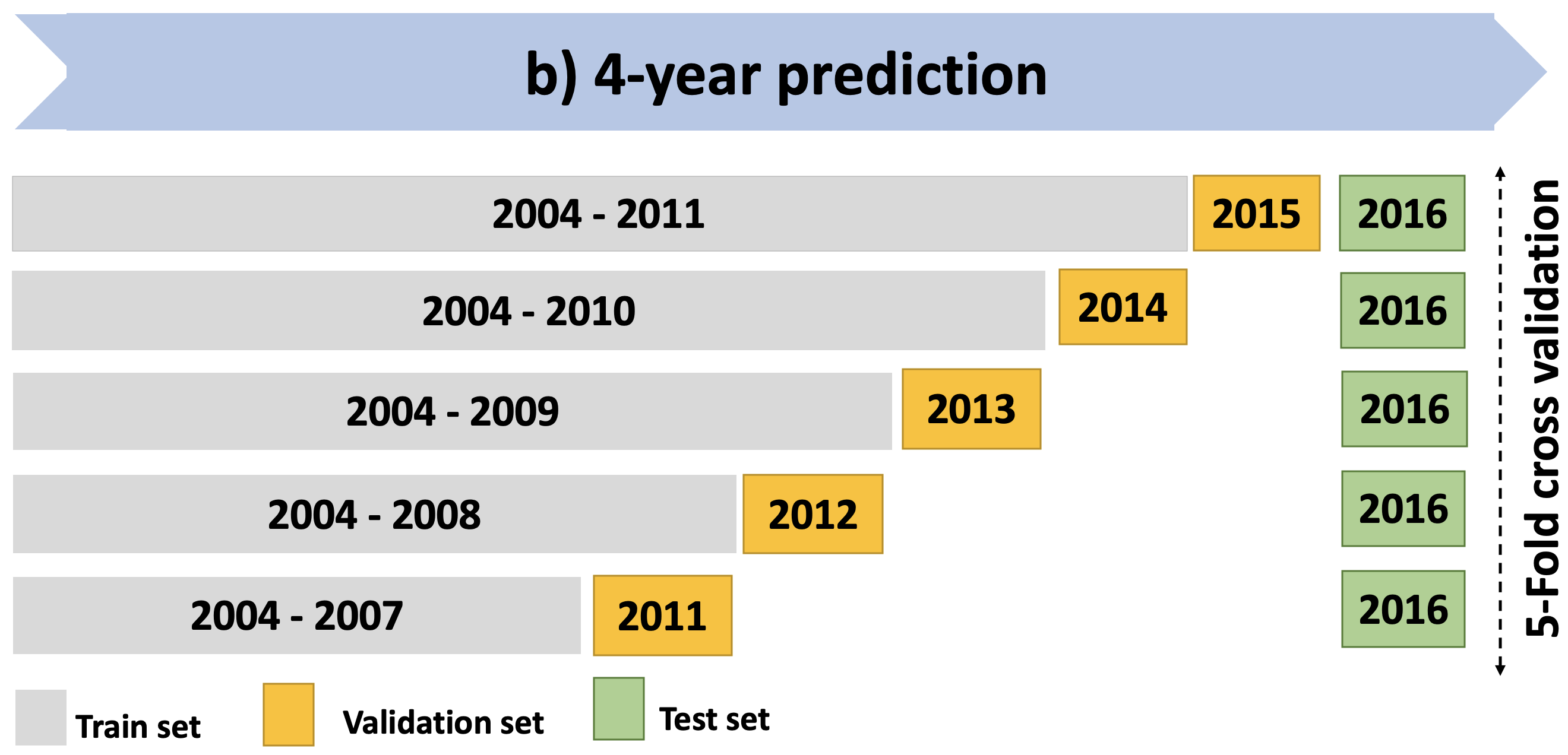}
 
\caption{Schematic of 5-fold cross-validation for a) 1-year prediction, and b) 4-year prediction.}

\label{fig:CV}
\end{figure}

To validate model performance, we employ a cross-validation type procedure by splitting the training dataset such that the information of the last year is taken as the new validation (dev) dataset and the remaining data as the training set. Figure \ref{fig:CV} shows a schematic of the cross-validation in which the train, validation and test datasets are splitted  for 1- and 4-year predictions. We perform the feature normalization and SMOTE over-sampling steps, which we have discussed in Section \ref{DataPre}, on the training dataset and transform the validation datasets according to the performed feature normalization. After retaining the evaluation scores, we discard the model, remove the validation data and repeat the same procedure on the remaining dataset. We repeat this procedure five times (5-fold)  to obtain five sets of evaluation scores for different validation years (2014-2018 for 1-year and 2011-2015 for 4-year). The final cross-validated scores are obtained by averaging these five sets of scores. 


\begin{figure}[!htp]
  \centering
  \includegraphics[width=0.49\textwidth]{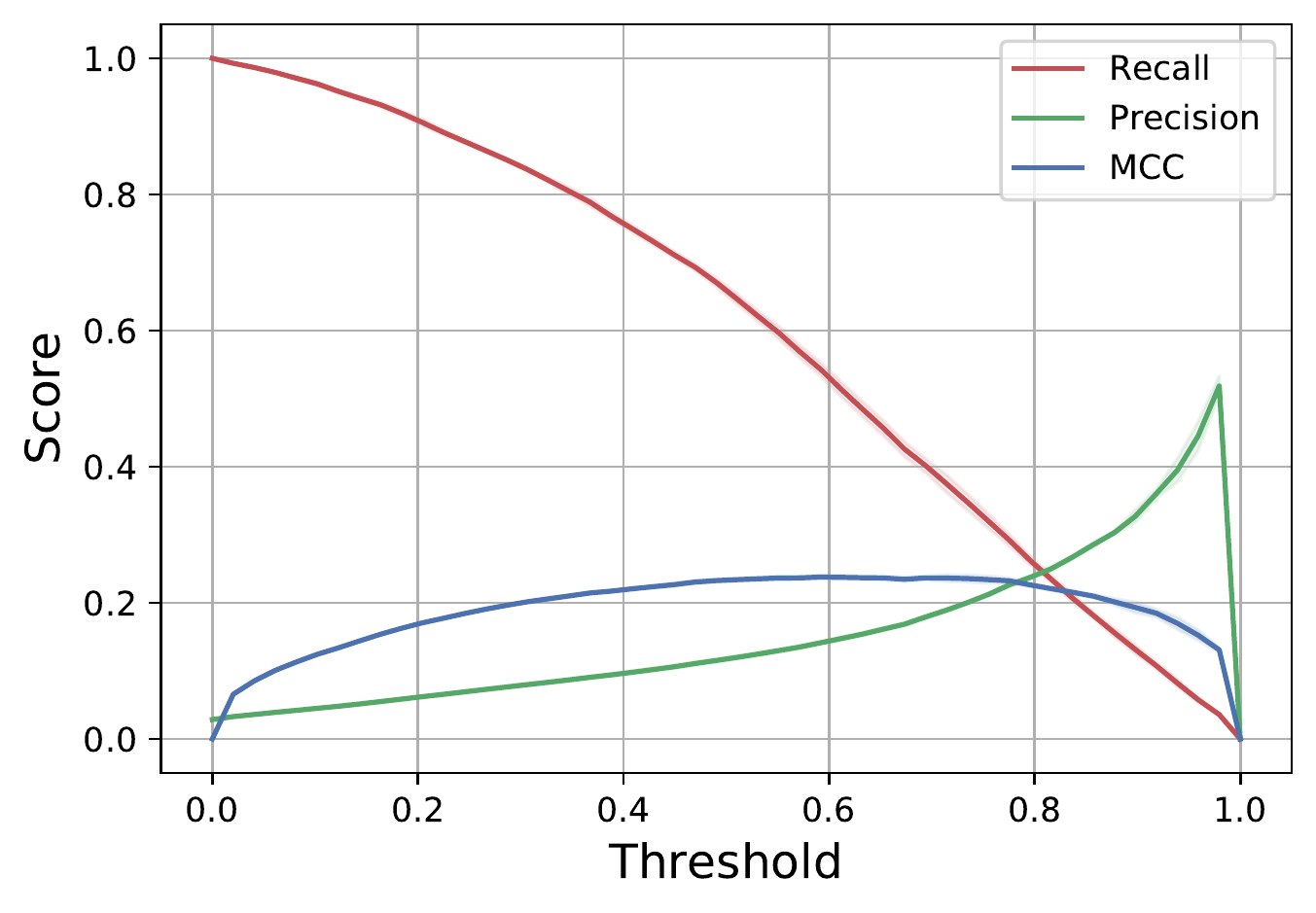}
  \includegraphics[width=0.49\textwidth]{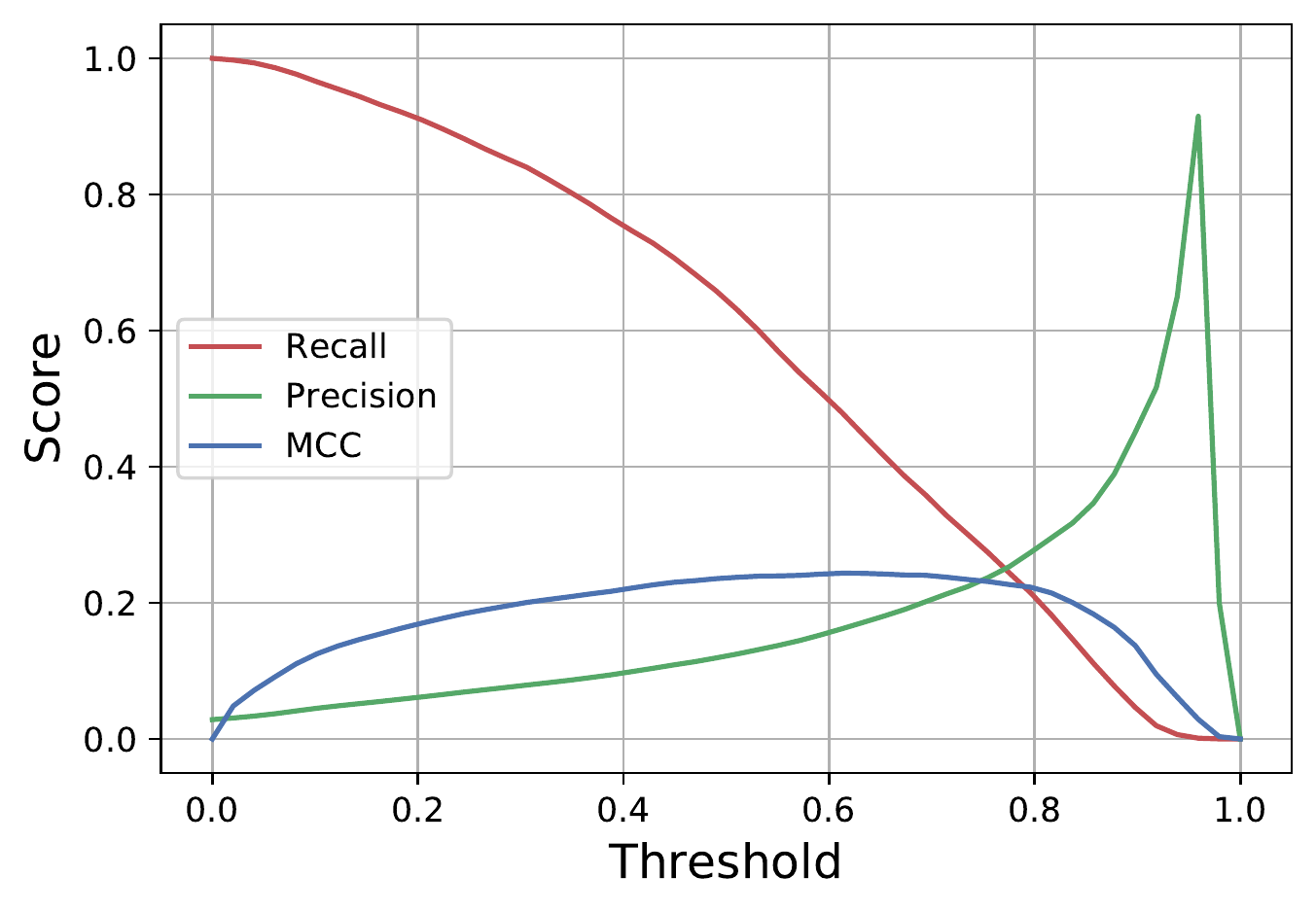} 
\caption{Cross-validation scores for precision, recall and MCC as a function of the classification threshold (decision threshold) using (left) Logit and (right) XGBoost classifiers. The solid line and shaded region represent the mean and standard deviation of the scores, respectively.}
\label{fig:CVscores}
\end{figure}

Having prepared the datasets and set the metrics, we employ two ML classifiers: (1) a Logistic Regression (Logit) classifier with L1 regularization to prevent over-fitting, and (2) an \textit{Extreme Gradient Boosting (XGBoost)} classifier which is an ensemble ML algorithm that forms a strong predictive model by combining multiple weak prediction models – often decision trees \citep{Chen2018}. Figure \ref{fig:CVscores} presents the cross-validation scores obtained from the Logit and XGBoost classifiers as a function of the classification threshold for the 4-year prediction. As expected, when the threshold approaches zero, the recall score tends to 1 and precision approaches to zero. Inversely, by increasing the threshold, the recall score decreases and precision increases up to a threshold value of around 0.95 for both classifiers. After identifying the threshold, where MCC is maximum, we perform a hyperparameter tuning to increase the MCC score considering several important parameters of each classifier as hyperparameters. Then we obtain unbiased scores using the trained model with the optimal hyperparameters and the test dataset of 2016. Table \ref{TableScore} presents these scores for each classifier. The highest MCC score of 0.243 at a threshold of around 0.6 belongs to the XGBoost classifier, highlighting the better performance of this model to predict pipe failures with respect to the Logit classifier. These results also indicate predictive accuracy and AUC higher than 80\% for both Logit and XGBoost classifiers.

\begin{figure}[!htp]
  \centering
  \includegraphics[width=0.7\textwidth]{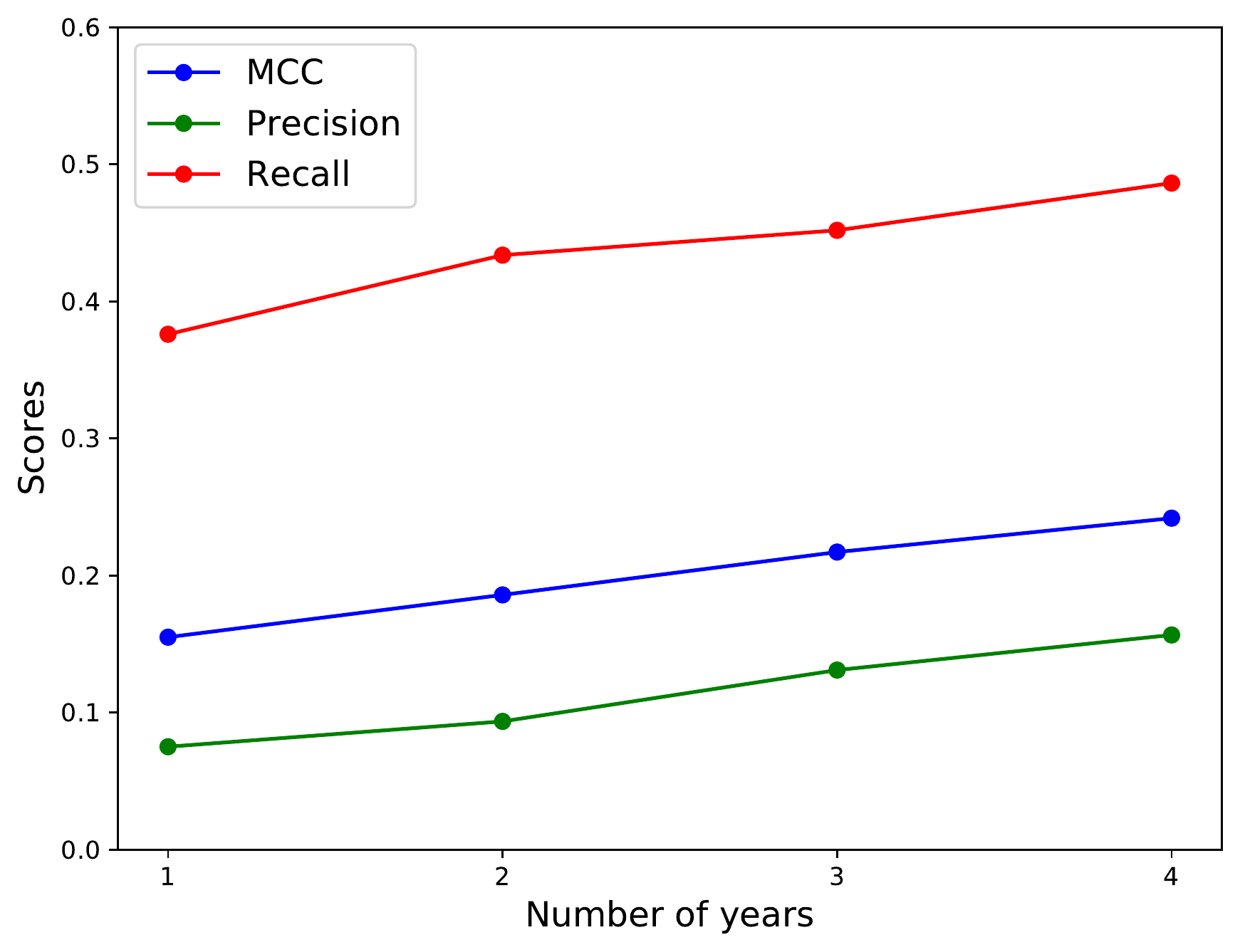} 
\caption{MCC, Precision and recall as a function of the number of years for failure prediction.}
\label{fig:Scores_Change}
\end{figure}

The above procedure is repeated for one-, two-, and three-year failure predictions, for which the target feature \textit{Failure} and the cross-validation procedure is adapted accordingly. Figure \ref{fig:Scores_Change} shows the MCC, precision and recall for these predictions. It is clear that the scores increases for longer period predictions, highlighting the effect of re-balancing the target feature and the SMOTE over-sampling.    

\begin{table}[!ht]
\renewcommand{\arraystretch}{1.50}
\centering
\begin{tabular}{p{20mm}|p{20mm} |p{20mm}}
\hline
\bfseries Score & \bfseries Logit  & \bfseries XGBoost\\
\hline
Threshold & 0.632 & 0.612 \\
MCC & 0.235 & 0.243 \\
Precision & 0.147 & 0.156 \\
Recall & 0.490 & 0.486 \\
Accuracy & 0.915 & 0.920 \\
f1 & 0.226 &  0.236\\
AUC & 0.854 & 0.859 \\
\hline
\end{tabular}
\caption{Performance metrics of the Logit and XGBoost classifiers on the test dataset (2016) for the 4-year prediction after hyperparameter tuning. The classification threshold is chosen where the mean of the cross-validated MCCs is maximum, i.e maximum of MCC for each classifier in Fig. \ref{fig:CVscores}.}
\label{TableScore}
\end{table}

\subsubsection{Predictor Importance}

Each feature or predictor contributes differently to the final prediction in both Logit and XGBoost models. We are interested in understanding how the models use the predictors in order to gain domain knowledge. 

Each method used offers different levels and forms of interpretation. For example, in the Logit model, each feature is associated to a coefficient that gives it weight in the final formula. If we normalize all the features to the same range, then we can compare directly the coefficients to extract the relative importance of each feature--although in this way we loose the units associated to each coefficient, so that to answer questions like what is the increase in probability by a 10 year increase in age we must first de-normalize the features and coefficients.

\begin{figure}[!htp]
  \centering
   \includegraphics[width=0.9\textwidth, angle =0]{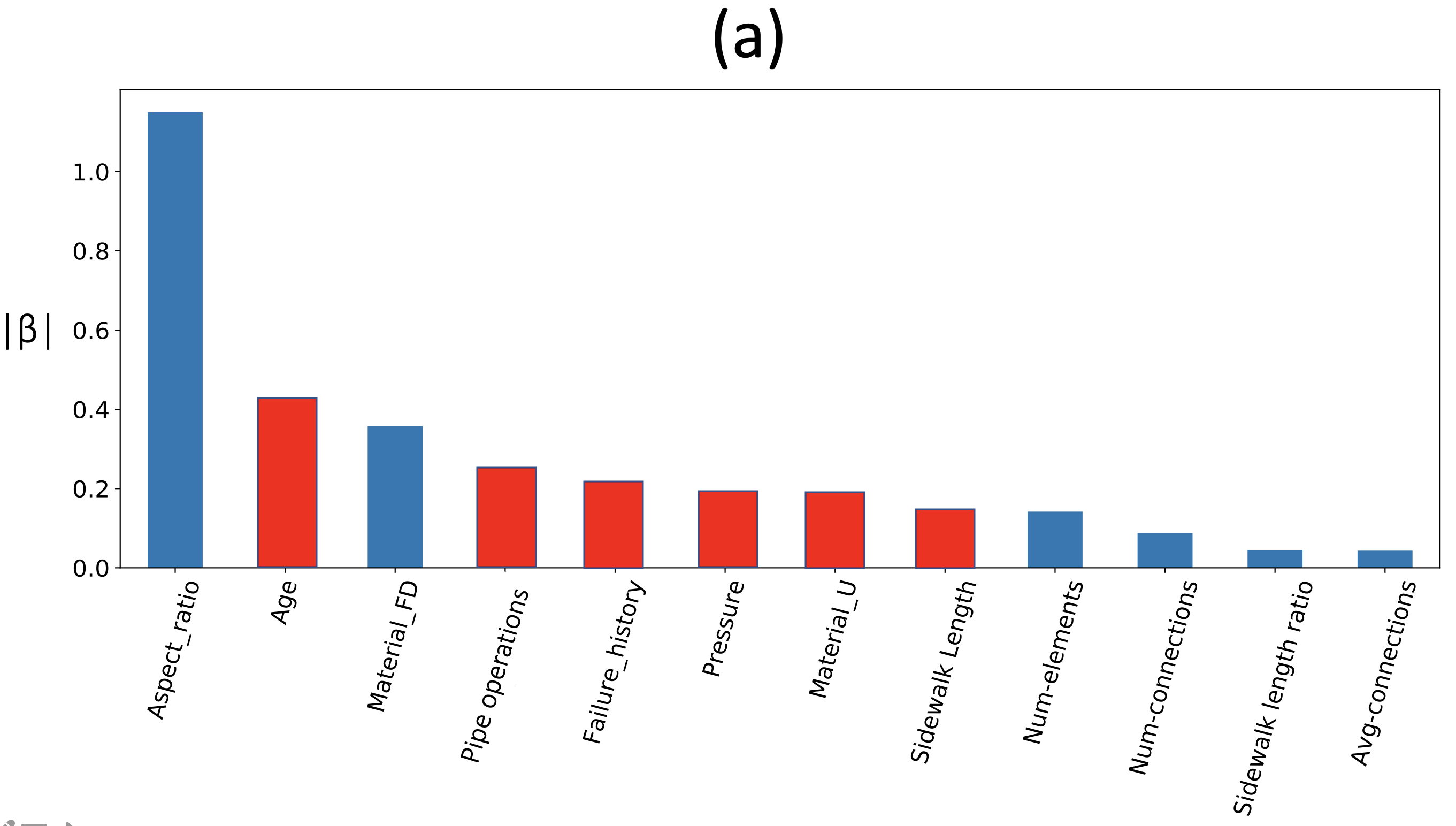}
   \includegraphics[width=0.8\textwidth, angle =0]{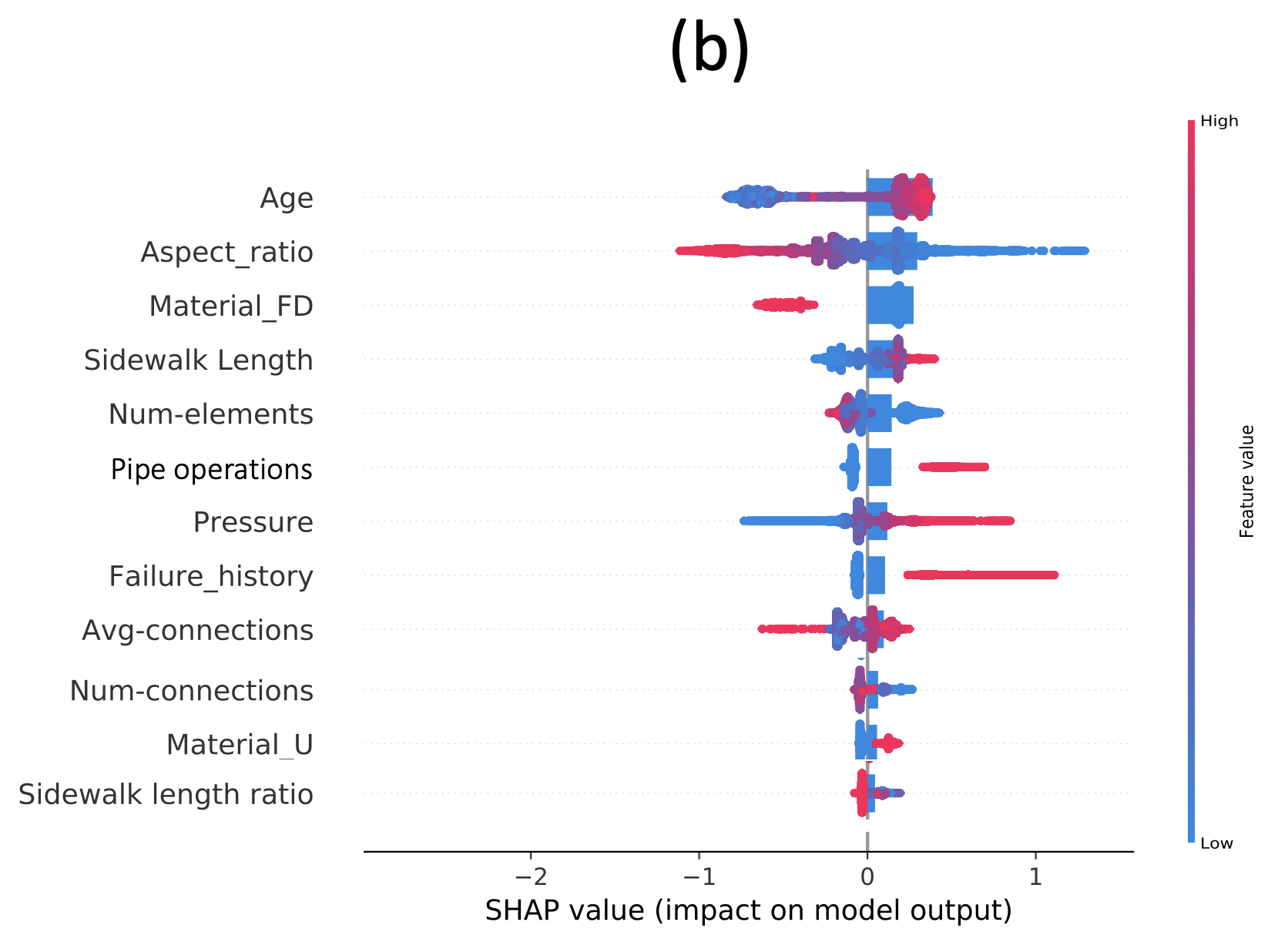}
\caption{Predictor importance as indicated by (a) Absolute value of Beta coefficients (feature weights) in the Logit model. Positive and negative values of Beta are highlighted in red and blue, respectively. (b) SHAP values in the XGBoost model. Features pushing the prediction higher are shown in red, those pushing the prediction lower are in blue.}
\label{fig:FeatureImp}
\end{figure}

In Figure \ref{fig:FeatureImp}(a) we plot the weights of each feature (its $\beta$ coefficient) for the optimized Logit model. As we can see, \textit{Aspect Ratio} (diameter divided by length) has the largest influence in the model. Its negative sign indicates that for higher pipe aspect ratios the probability of failure is lower--although we notice again that aspect ratio reflects a combination of geometric properties with the inventory practices used to assign length to each pipe ID. Pipes made of \textit{Material-FD} (ductile iron) also show the same trend, highlighting the resistance of this material against failures. Other relevant features are the \textit{Age} of the pipe, nearby failures (\textit{Pipe operations}), and the pipe's own \textit{Failure history}. For all these features, higher values features increase the probability of pipe failure. 

The XGBoost model needs more work to extract this kind of information, as it is composed of large ensembles of decision trees that might contradict each other. Popular feature attribution techniques for tree ensemble methods (such as counting the number of times a feature is taken into account) are inconsistent \citep{Lundberg2018} and led to the use of SHapley Additive exPlanation (SHAP) values, a technique that uses fast exact tree solutions based on a unification of ideas from game theory. The SHAP values consistently attribute feature importance, better align with human intuition, and better recover prominent features. However, we find that in general the analysis of SHAP values agrees with the interpretation of the Logit model coefficients. The SHAP values for our XGBoost model, shown in Fig. \ref{fig:FeatureImp}(b), indicate that \textit{Age} and \textit{Aspect Ratio} are the most important predictors. As with the Logit model, pipes of \textit{Material-FD} and with larger values of \textit{Aspect Ratio} decrease the failure probability. \textit{Pipe operations} and \textit{Failure history} are also among the influential features correlating positively with  failure probability. Notice that the sign of the Logit model coefficients (Fig. \ref{fig:FeatureImp}(a)) are in general agreement with the impact of features on the SHAP values (Fig. \ref{fig:FeatureImp}(b)).        

\subsection{Survival Analysis }

\subsubsection{ Cox proportional hazards model}
The Cox's Proportional Hazards Model is defined based on a hazard function, which is the product of a completely unspecified baseline hazard function and a parametric function of predictors. The hazard function at time $x$ for a pipe with predictor vector $z$ is stated as \citep{Cox1972}:

\begin{equation}
h ( x | \mathbf { z } ) = h _ { 0 } ( x ) \exp \left( \mathbf { z } \boldsymbol { \beta } \right) = h _ { 0 } ( x ) \exp \left( \sum _ { i = 1 } ^ { p } \beta _ { i } z _ { i } \right)
\label{COX}
\end{equation}
where $h _ { 0 } ( x )$ is the baseline hazard function, $\boldsymbol { \beta } = \left( \beta _ { 1 } , \beta _ { 2 } \ldots \beta _ { p } \right)$ is a parameter vector and $\boldsymbol { z }$ is a vector of covariates $\left( \mathbf { z } = \left[ \mathbf { z } _ { 1 } , . . \mathbf { z } _ { \mathrm { p } } \right] \right)$. The baseline hazard function is the hazard function corresponding to the situation when all of the covariates in the model are equal to zero. To avoid overfitting, we use CoxNet, a version of Cox model with an elastic net regularization, which is a penalized likelihood method that combines $L_{1}$ (lasso) and $L_{2}$ (ridge) regularizations \citep{Young2007,Noah2011}. 

To prepare the dataset for the CoxNet model, we keep the last record of each pipe before failure or the record in 2019 for pipes without any failure (censored). In this analysis, the target feature is \textit{Age}, which is considered as the time from the installation year to the time of failure or the end of the study in the case of no failure. Instead of pipe age, the installation year (\textit{Install Date}) is retrieved from the original data as an independent feature. In order to avoid data leakage, we remove the features \textit{Avg-connections}, \textit{Avg-connections ratio}, and \textit{Avg-elements} which could bias the model towards the target feature \textit{Age}. 

\begin{figure}[!t]
  \centering
  \includegraphics[width=0.9\textwidth]{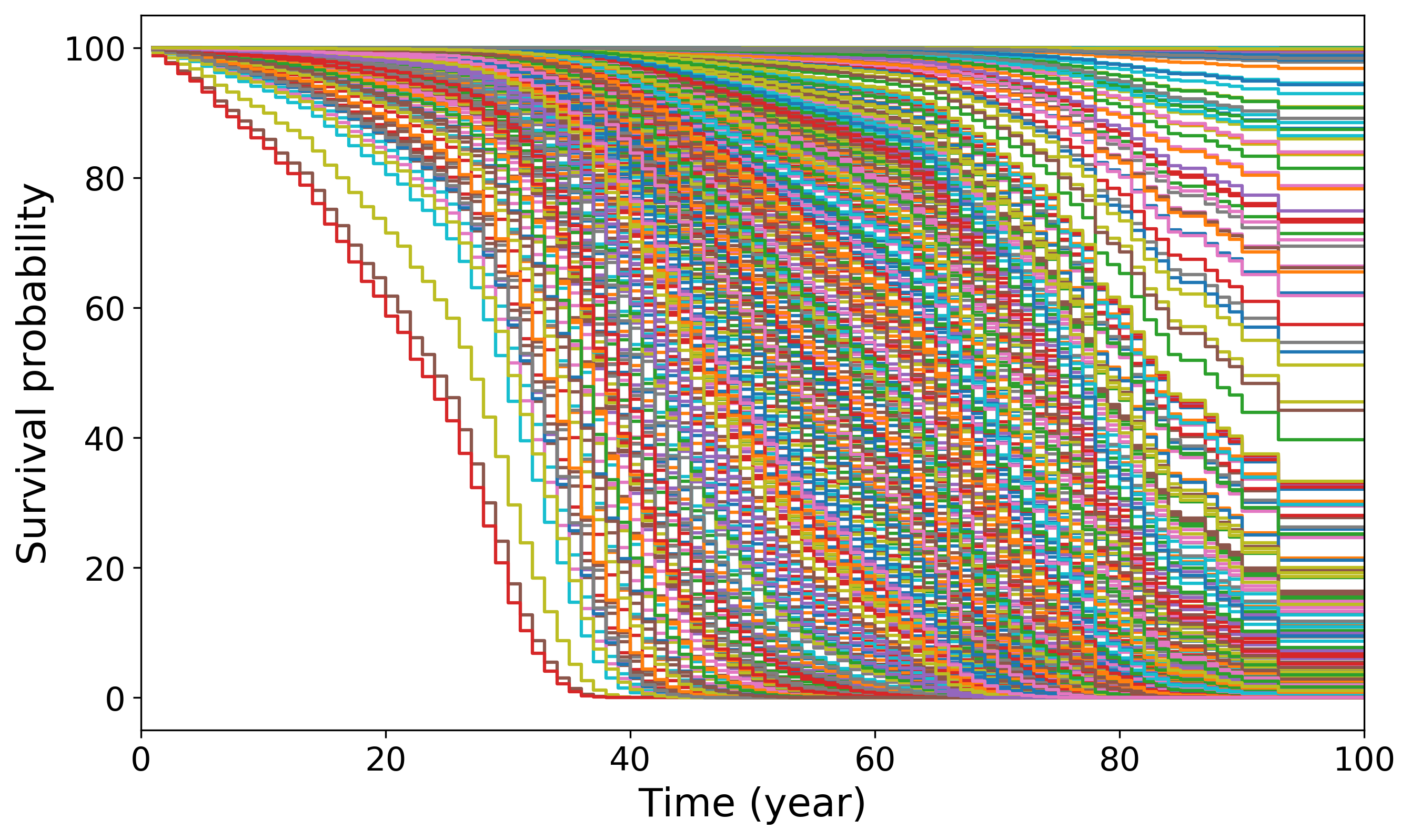}
\caption{Survival probability of a random sample of 500 water pipes as a function of time (in year).}
\label{fig:Survival}
\end{figure}

After splitting the dataset into train (70\%) and test (30\%) sets, we train the CoxNet model and measure the quality of the fit through the weighted Brier score (BS), which is a modification of the Brier score to prevent a bias towards censored data in survival models \citep{Thomas2006,Young2010}. This is particularly important to reflect a fair performance of our model since the majority of observations in our imbalanced dataset are censored. The weighted Brier score is stated as:

\begin{equation}
\mathrm{BS}=\frac{1}{NT}\sum_{i=1}^{N} \sum_{j=1}^{T} w_{i}(t_j)\left[p_{i}(t_j)-d_{i}(t_j)\right]^{2}
\end{equation}
where $p_{i}$ is the predicted probability of survival for  the $i$-th pipe at time $t$ of analysis, $d_{i}$ is the actual outcome of the event (0 for failure and 1 for non-failure), $N$ is the total number of pipes in the dataset, $T$ is the prediction window, and $w_{i}$ is the weight assigned to each pipe at time $t$, calculated using the inverse probability of censoring weights method \citep{Graf1999,Gerds2013}.  A weighted Brier score of around 0.016 is obtained on the test set with a prediction window of $T$ = 100 years, an acceptable score for the model. Note that the Brier score is always a number between 0 and 1, with 0 being the best possible value. In  Fig. \ref{fig:Survival}, we plot the survival curves of 500 pipes in the training set. These pipes show a wide variety of survival probability curves. The minimum predicted survival time is around 38 years and there are many pipes that could reach up to 100 years with a survival probability of over 50\%. 

\begin{figure}[!htp]
  \centering
  \includegraphics[width=1.0\textwidth]{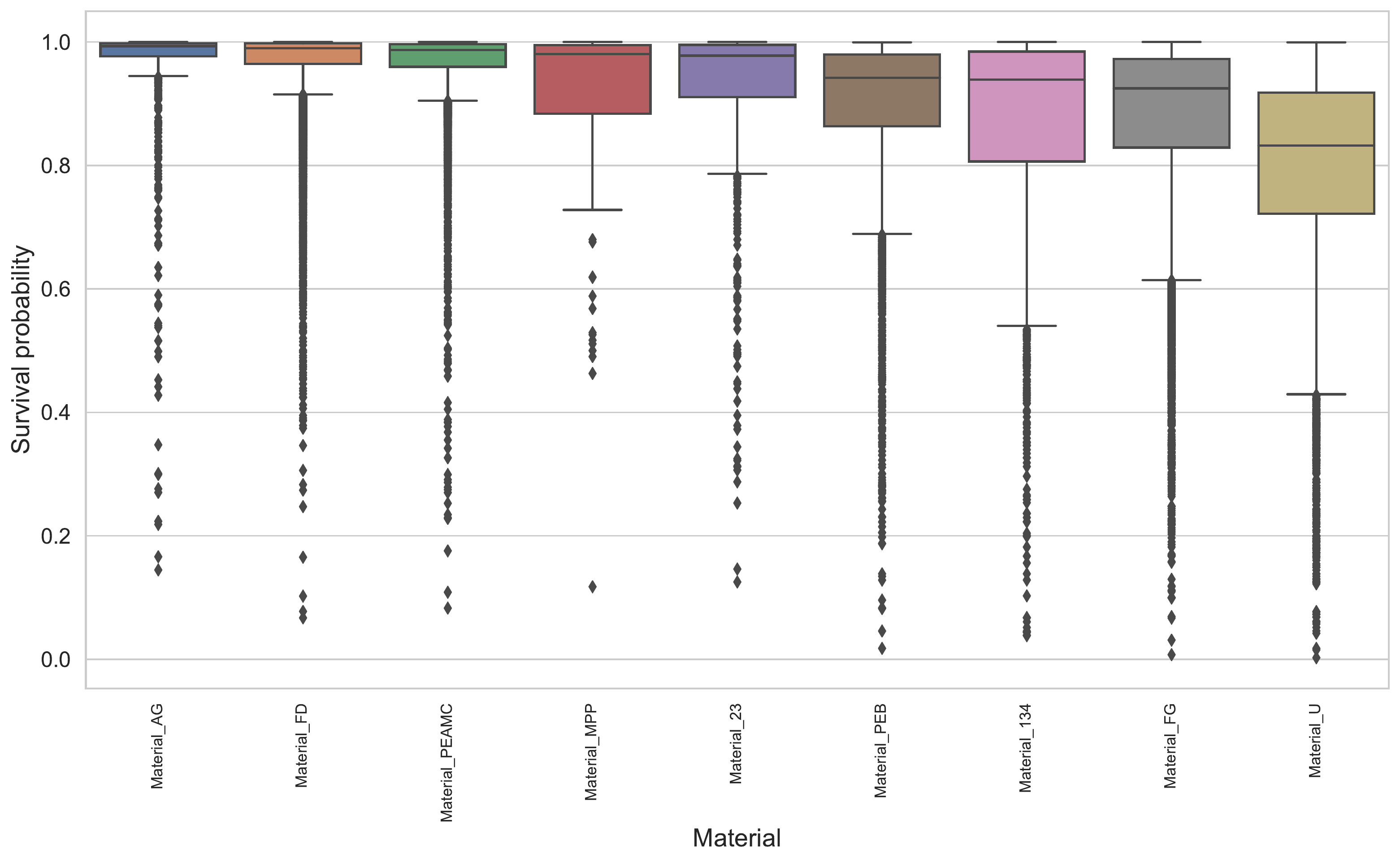}
\caption{Boxplot of survival probability of water pipes in service for each material category. The boxes are sorted in descending order from left to right according to the median of survival probability in each category. The survival probability is obtained for each pipe in two years (2022) }
\label{fig:SurvivalMaterial}
\end{figure}

To provide additional insight from the distribution of survival probability, we also plot this distribution for each material category in Fig. \ref{fig:SurvivalMaterial}. It is clear that among the most widely used materials for fabricating the pipes (FD and FG), the ductile iron (FD) pipes show a higher range of survival probability. This result is supported by the predictions of the Logit and XGBoost models which show the high importance of the predictor \textit{Material-FD}, see Fig. \ref{fig:FeatureImp}. Polyethylene (PEAMC) pipes show a similar range to that of the FD pipes, whereas a median survival probability of more than 95\% is observed for steel (AG) pipes.

\subsubsection{Predictor Importance }

\begin{figure}[!htp]
  \centering
  \includegraphics[width=1\textwidth]{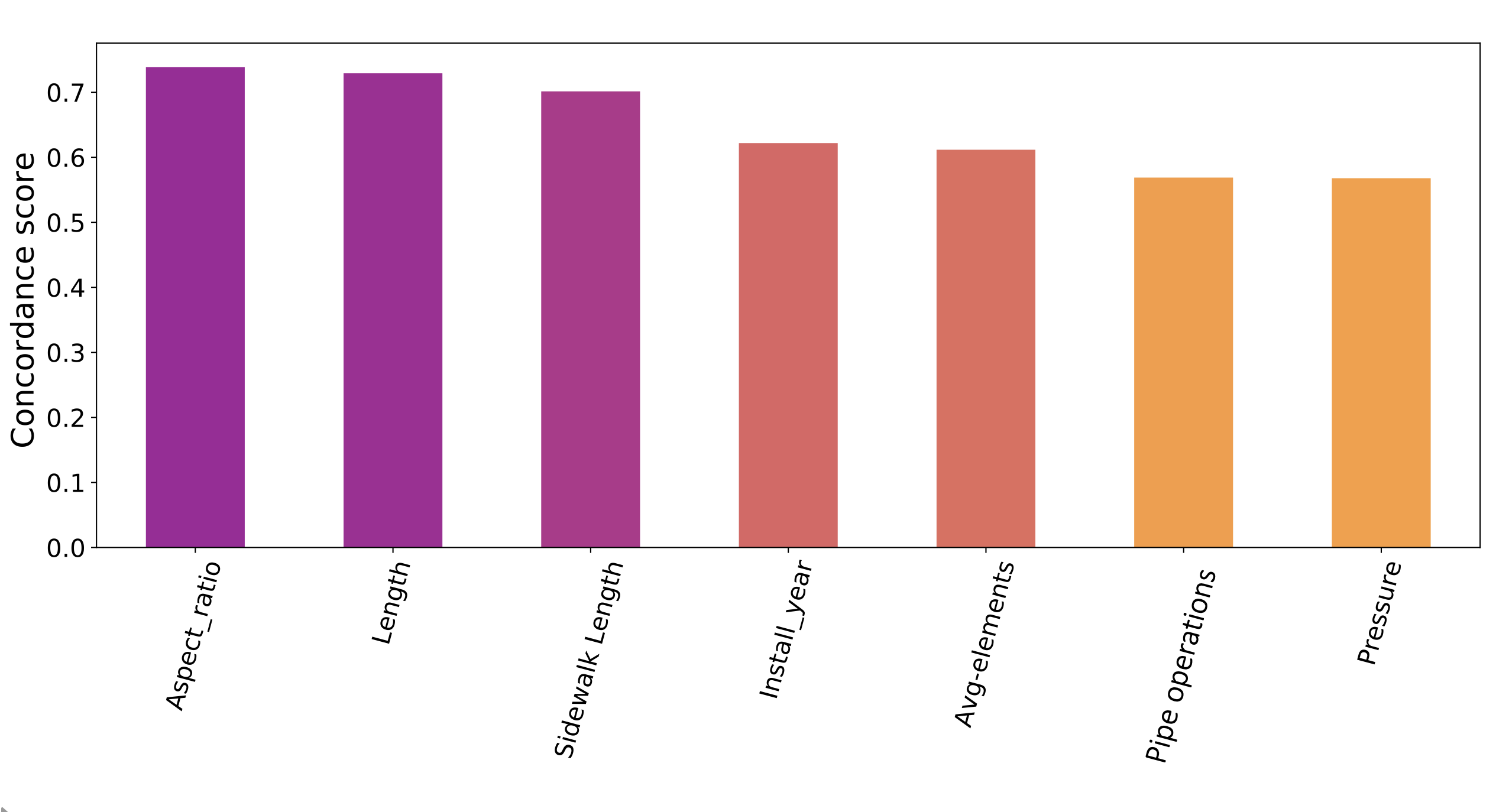}
\caption{Concordance score of the survival model, trained with each individual predictor. }
\label{fig:COXCI}
\end{figure}

To highlight the important features in predicting survival probability, we have obtained the concordance score (also known as the c-index) of the survival model \citep{Vikas2007}. This score is commonly used for performance evaluation of survival models and is a measure of the predictive accuracy of the fitted model onto the training dataset, analogous to the coefficient of determination (R$^2$) in linear models. The concordance score is calculated by training the survival model with each individual predictors in the dataset. Figure \ref{fig:COXCI} shows this score for important predictors with a concordance score of more than 0.55, a baseline  score for comparison of predictors in Cox survival models \citep{Therneau2015}. It is interesting to observe that the model that only includes \textit{Aspect Ratio} leads to the highest score, in agreement with the short term failure models discussed above. 
The \textit{Install-year} feature also appears to have a high impact, in agreement with the high importance of the predictor \textit{Age} given by the LR and XGBoost classifiers.  

\section {Conclusions}
We have developed a statistical and machine learning framework for predicting pipe failure and survival of Barcelona water distribution network. After performing data preprocessing and introducing new features to the original dataset, we have employed two classification methods for the prediction of pipe failure in the upcoming one to four years. These methods are Logistic Regression (Logit) and Extreme Gradient Boosting (XGBoost) classifiers. Before training these models, we have identified that the given dataset is imbalanced, a particular case where 99\% of data belongs to the negative class (pipes without failure). Since training the classifiers with this imbalanced dataset results in poor performance of the models, we have employed a re-sampling method called Synthetic Minority Over-sampling Technique (SMOTE) to balance the data with synthetic examples. We have used the re-sampled dataset to train the Logit and XGBoost models following a cross-validation scheme. We have obtained cross-validated scores considering the Matthews Correlation Coefficient (MCC) as a proper performance metrics for imbalanced datasets. A maximum MCC of around 0.24 has been obtained using the XGBoost classifier.  

We have also performed a predictor importance analysis to highlight the important features in predicting pipe failure. By obtaining standardized weights from the Logit model, we have concluded that pipe geometry, age, and the number of previous failures and operations are the most critical risk factors. This conclusion is supported by another predictor importance analysis that we have done using SHAP (SHapley Additive exPlanation), a recent feature attribution technique for tree ensemble methods such as the XGBoost model. This analysis has again shown that pipe geometry and age are the most important predictors.  

Finally, to provide a broader perspective of pipe failure, we have performed a survival analysis using Cox's Proportional Hazards Model with an elastic-net regularization to avoid overfitting. Moreover, the survival probability of each individual pipe has been obtained, and different materials have been compared based on their survival probability distributions. This could suggest reliable pipe materials and help utility managers to prioritise the most critical pipes for rehabilitation. The concordance score given by the survival model for each individual feature confirms that pipe geometry is one of the most critical risk factors.

%
%
\bibliography{biblio}

\end{document}